\documentclass[]{aa}  

\usepackage{graphicx}  

\usepackage[varg]{txfonts}
\usepackage{hyperref}
\hypersetup{
    colorlinks=true,
    linkcolor=blue,
    filecolor=magenta,      
    urlcolor=cyan,
   citecolor=blue,
}
\newcommand{\BLOS}{\ensuremath{B_\mathrm{LOS}}}

\newcommand{\vLOS}{\ensuremath{v_\mathrm{LOS}}}
\newcommand{\vPOS}{\ensuremath{v_\mathrm{POS}}}

\defcitealias{2020kri}{Paper I}
\begin{document}

   \title{Magnetic field inference in active region coronal loops using coronal rain clumps}

   \author{M. Kriginsky\inst{1,2} \and R. Oliver\inst{1,2}
 \and P. Antolin \inst{3} 
\and D. Kuridze\inst{4,5}
 \and N. Freij\inst{1}
          }

   \institute{Departament de F\'\i sica, Universitat de les Illes Balears, E-07122 Palma de Mallorca, Spain
         \and
          Institute of Applied Computing \& Community Code (IAC3), UIB, Spain
          \and
            Department of Mathematics, Physics and Electrical Engineering, Northumbria University, Newcastle Upon Tyne, NE1 8ST, UK
           \and
            Department of Physics, Aberystwyth University, Ceredigion, SY23 3BZ, UK 
          \and
            Abastumani Astrophysical Observatory, Mount Kanobili, 0301, Abastumani, Georgia
                      }

   \date{Received ; accepted }

 
  \abstract
   {}
   {We aim to infer information about the magnetic field in the low solar corona from coronal rain clumps using high-resolution spectropolarimetric observations in the \ion{Ca}{II} 8542 \AA\ line obtained with the Swedish 1-m Solar Telescope. }
   {The weak-field approximation (WFA) provides a simple tool to obtain the line-of-sight component of the magnetic field from spectropolarimetric observations. We adapted a method developed in a previous paper in order to assess the different conditions that must be satisfied in order to properly use the WFA for the data at hand. We also made use of velocity measurements in order to estimate the plane-of-the-sky magnetic field component, so that the magnetic field vector could be inferred.}
   {We have inferred the magnetic field vector from a data set totalling 100 spectral scans in the \ion{Ca}{II} 8542 \AA\ line, containing an off-limb view of the lower portion of catastrophically cooled coronal loops in an active region. Our results, albeit limited by the cadence and signal-to-noise ratio of the data, suggest that magnetic field strengths of hundreds of Gauss, even reaching up to 1000 G, are omnipresent at coronal heights below 9 Mm from the visible limb. Our results are also compatible with the presence of larger magnetic field values such as those reported by previous works. However, for large magnetic fields, the Doppler width from coronal rain is not that much larger than the Zeeman width, thwarting the application of the WFA. Furthermore, we have determined the temperature, $T$, and microturbulent velocity, $\xi$, of coronal rain clumps and off-limb spicules present in the same data set, and we have found that the former ones have narrower $T$ and $\xi$ distributions, their average temperature is similar, and coronal rain has microturbulent velocities smaller than those of spicules.} 
   {}


   \titlerunning{Using coronal rain clumps for the inference of coronal magnetic fields}
   \authorrunning{M. Kriginsky et al.}
   \maketitle
   
%

\section{Introduction}

As the history of the observations of prominences unfolded, they were classified into different categories of a similar phenomenon. One class of prominence that could be seen in the surroundings of active regions was named the Class IV prominence by  \citet{1943ApJ....98....6P}\footnote{Pettit's classes IIIa and IIIb may also correspond to coronal rain material, but the evidence presented by this author is inconclusive.}. This type of prominence contained material that was falling in the shape of streamers of coronal material or, as \citet{1943ApJ....98....6P} called them, 'coronals'. Coronals can be observed in $\mathrm{H\alpha}$ and \ion{Ca}{II} H and K lines falling into the upper chromosphere with velocities of $100-200$ km $ \mathrm{s}^{-1}$, several of them being along a common path. One of the first works to coin the term 'coronal rain' to characterise material falling into the chromosphere was the classification of prominences developed by \citet{MenzelEvans1953}, although they interpreted that the precipitating material originated from condensations of coronal plasma and that the phenomenon occurred outside of active regions. Only recently has it been shown that coronal rain is in fact a phenomenon that should be differentiated from prominences due to the very specific conditions under which it occurs \citep[e.g.][]{2010ApJ...716..154A}. 

Some numerical simulations show that under the influence of a steady, uniform heating, coronal loops are found to be in a state of thermal equilibrium \citep[see, for example,] [and the references therein]{2013ApJ...773...94M}. This equilibrium is either static or consists of a steady flow of material from one end of the loop to the other. However, there are certain situations for which no possible equilibrium can be achieved, even in the presence of a constant or high frequency impulsive heating. This state is called thermal nonequilibrium \citep[TNE;][]{1991ApJ...378..372A,2019ApJ...884...68K}. The heating of the loop in these situations is mainly concentrated at the footpoints, and the dynamics of the plasma are constantly evolving in response to the energy imbalance present in the loop due to the inefficiency of thermal conduction to compensate for the radiative losses in the uppermost part of the loop.  This imbalance leads the temperature and density to reach critical values, a phenomenon known as 'catastrophic cooling' \citep{1999ApJ...512..985A,1991ApJ...378..372A}. The radiative losses increase as the density grows and the temperature decreases. This evolution can sometimes lead to the formation of cold, high-density condensations of plasma inside coronal loops because, locally, thermal instability can take place and drive the local accretion of plasma into high-density clumps. This condensed material can later fall back along the loop to the chromosphere in the form of coronal rain \citep{2001SoPh..198..325S, 2004A&A...424..289M,2010ApJ...716..154A}.  If the conditions are maintained over a timescale of the order of the radiative timescale, the coronal rain events can happen periodically with a timescale that depends on several factors, including the specific geometry of the loop or the nature of the heating \citep{2017ApJ...835..272F,2018ApJ...855...52F}. The repeating cycle corresponds to periods of evaporation of chromospheric material that accumulates within the loop and periods of drainage of part of the material inside the loop in the form of precipitation that falls along the magnetic field lines.  These condensation-evaporation cycles have their extreme ultraviolet (EUV) observational counterparts in the long period EUV pulsations, which have been described in recent studies \citep[see][]{2020A&A...634A..54P,2016ApJ...827..152A,2018ApJ...853..176A,2020A&A...633A..11F}.

Coronal rain can appear mainly in two situations \citep{2020PPCF...62a4016A}. It can take place during the gradual phase of a flare, within the flaring loops at the late stage of their cooling phase, or, more commonly, in active region loops. Another coronal rain scenario has been recently identified in the quiet Sun, associated with null point topologies high up in the corona, where magnetic field dips are naturally formed and where material can accumulate \citep{2019ApJ...874L..33M,2018ApJ...864L...4L}. The appearance of the clumps heavily depends on the spectral line used to observe them. They have a very wide temperature range, from a  few $10^3$ K up to several times $10^5$~K, so they can be seen both in chromospheric and transition region lines. When observed in chromospheric lines, coronal rain appears clumpy and is seen to have a smaller length (hundreds of kilometers) measured along the direction of the flow than when seen through transition region lines, for which much longer lengths (up to tens of megameters) are observed \citep{2010ApJ...716..154A,2015ApJ...806...81A,2012ApJ...745..152A}. However, thinner and more elongated clumps are also seen in lines such as $\mathrm{H\alpha}$, and we have also spotted several of them in our observations in the \ion{Ca}{II} 8542 $\AA$  line. 

The structure of coronal rain clumps is assumed to consist of a cool core, surrounded by a thin transition region of a higher temperature that envelops the clump and serves as a boundary between the condensed material and the corona \citep{2020PPCF...62a4016A}. Multiple individual coronal rain clumps can fall along the same coronal loop or bundle of loops at the same time, so the appearance of what can be perceived as one individual coronal rain clump can change with time given the possible superposition of multiple clumps when viewed with a coarse spatial resolution instrument. Instruments that provide higher spatial resolution, such as the Daniel K. Inouye Solar Telescope \citep[DKIST;][]{2020SoPh..295..172R}, will likely reveal that the widths of the clumps have a distribution of even smaller values than those measured with current instruments \citep{2014ApJ...797...36S,2015ApJ...806...81A}.

The plasma that constitutes coronal rain consists of both neutrals and ions, and we assume them to be in a state of strong coupling, which is predominantly achieved through collisions between the ions and the neutral particles \citep{2016ApJ...818..128O}. This means that even when observed through neutral lines such as  $\mathrm{H\alpha}$, the motion of coronal rain clumps can be safely assumed to take place along the magnetic field lines of the coronal loop. The implications of this fact are cardinal to the study of the morphology and structure of the magnetic field in coronal loops. Coronal rain has already been used to map the topology of the magnetic field of coronal loops \citep{2012ApJ...745..152A}. With spectropolarimetric observations of coronal rain clumps, it is also possible to set further constraints on the magnetic field strength of coronal loops.

Direct inferences of magnetic fields of coronal loops from spectropolarimetric observations are rather scarce. \citet{2016ApJ...833....5S} employed observations obtained using the Facility Infrared
Spectropolarimeter \citep[FIRS;][]{2010MmSAI..81..763J} at the 0.76 m Dunn Solar Telescope \citep[DST;][]{1964ApOpt...3.1353D,1991AdSpR..11e.139D} in the \ion{He}{I} 10830 $\AA$ line in order to study the magnetic field topology and structure of a coronal loop filled with cold material. The data had a spatial sampling of 0.29\arcsec~pixel$^{-1}$, which equalled the size of each scan step. A total of 218 steps were scanned in an observation which took 32 minutes to generate, resulting in a 60\arcsec $\times$70\arcsec field of view. The magnetic field vector along a coronal loop was inferred using the inversion code $\mathrm{HELIX}^{+}$ \citep[He-Line Information Extractor+;][]{2004A&A...414.1109L,2007AdSpR..39.1734L}. With the results from the first part of the inversion, accounting only for the Zeeman effect,  these authors were able to study the variation with respect to the height above the solar surface of the magnetic field component along the line of sight (\BLOS). Under the assumption that the magnetic field and the plasma flow are aligned with the orientation of the triangulated loop, the total magnetic field strength was calculated. We note that $B$ and \BLOS\hspace{1pt}strengths higher than 1000~G were inferred at a height of 9 Mm above the limb, with $B$ decreasing to only a few Gauss at coronal heights above 50 Mm. The full magnetic field strength was also inferred using the 'Hanle-slab' model of the $\mathrm{HELIX}^{+}$ code applied to the highly redshifted \ion{He}{I} component. These authors report that the only coherent results obtained using this method were for heights ranging between 14 and 50 Mm, because outside this range the strength of the linear polarisation signal was not sufficient to yield reliable results. The type of solutions that yielded a magnetic field aligned with the coronal loop geometry turned out to be those that were most similar to the results inferred in the first step of the analysis, using the triangulated loop to obtain the magnetic field strength.

\begin{figure*}
   \centering
   \includegraphics[width=9cm]{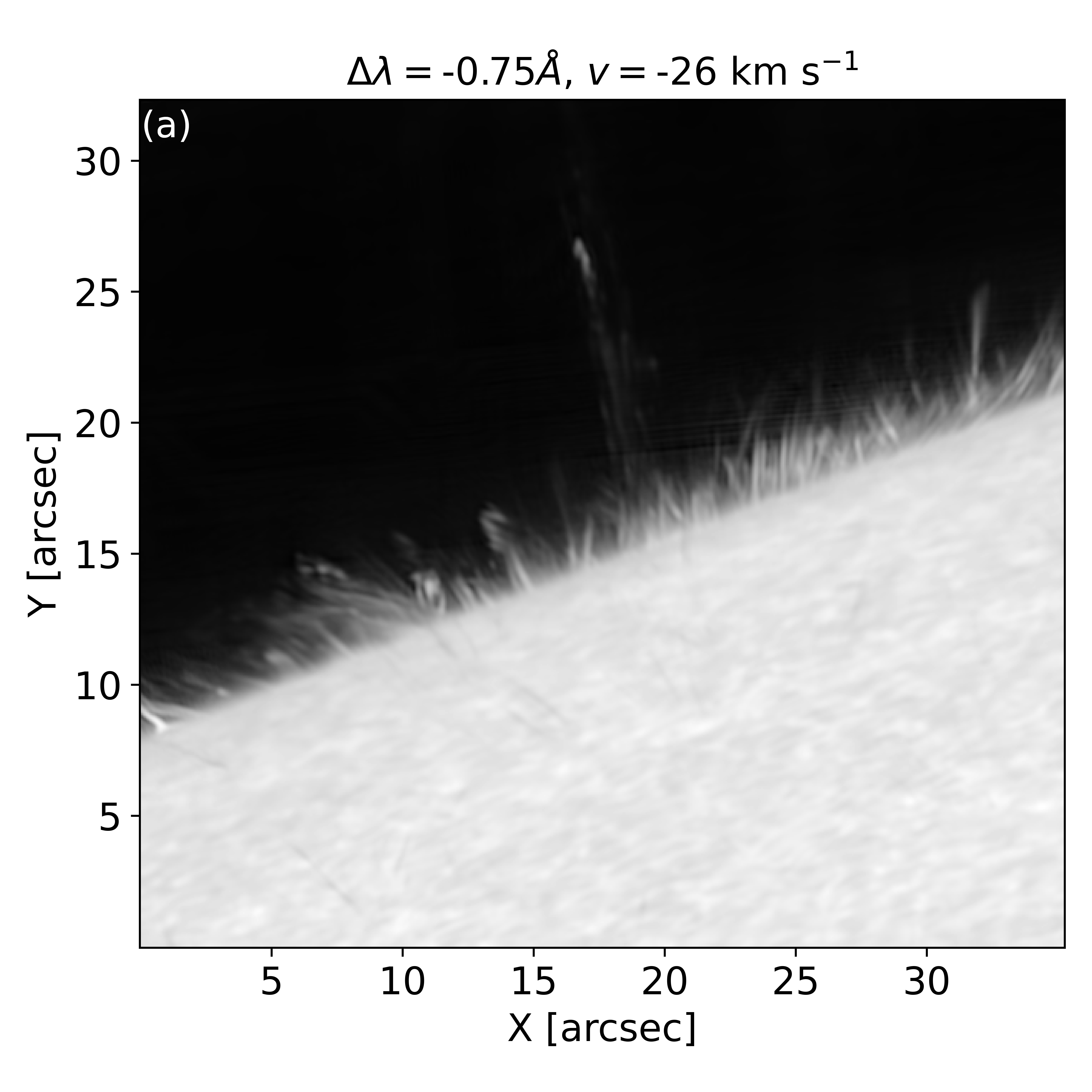}
    \includegraphics[width=9cm]{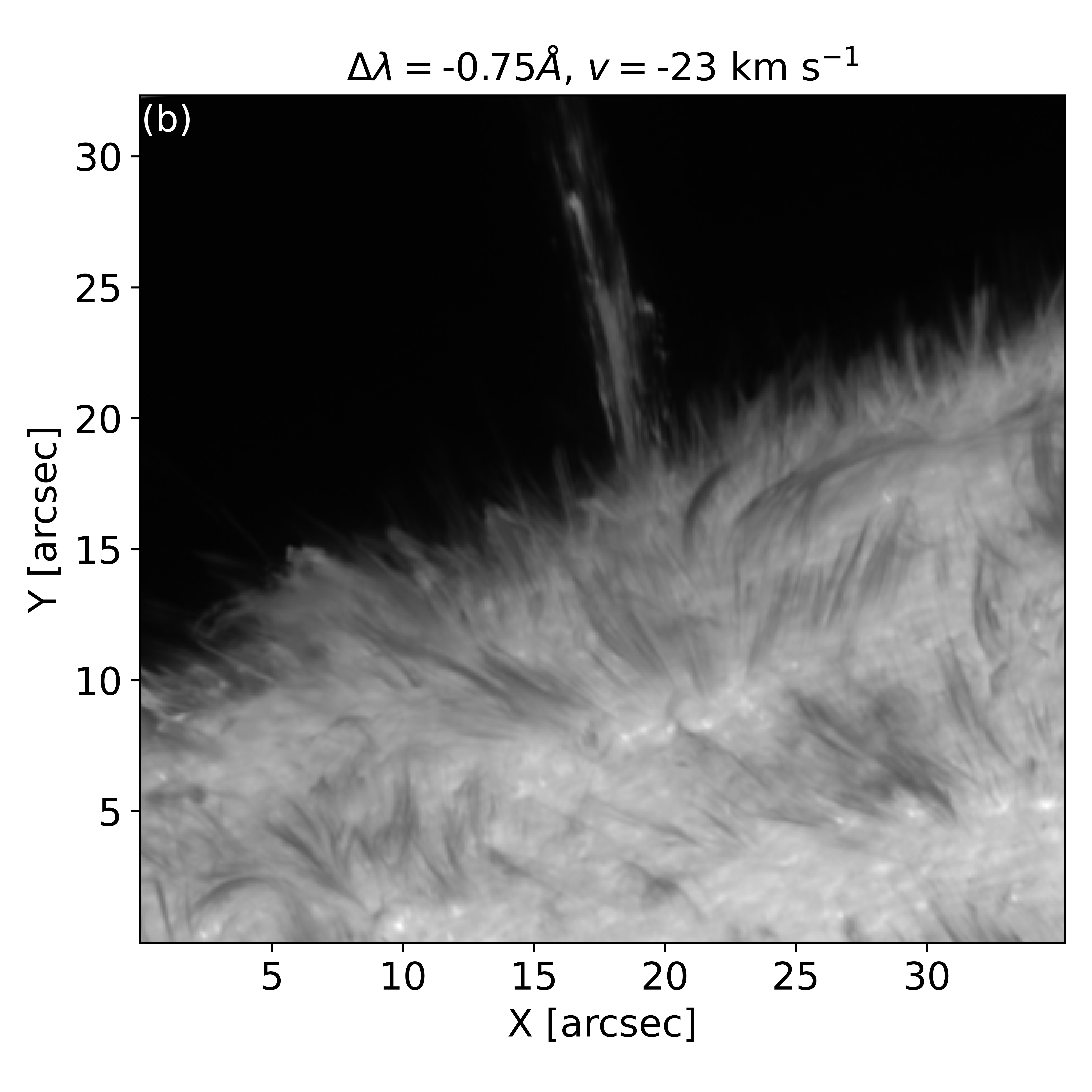}

   \caption{Evolution of coronal rain during the observations. (a) In the \ion{Ca}{II} 8542 \AA\ line at 16:40:55 UT and (b) in the H$\alpha$ at 16:40:41~UT. The logarithm of the intensity is shown in panel (a)  to enhance the appearance of the clumps, which otherwise would not be visible due to the much larger average intensity from the disc. An animation associated with the observations is available online, where the \ion{Ca}{II} 8542 \AA\ intensity shown in a copper coloured table is superposed on the H$\alpha$ intensity displayed in a grey coloured table.}
              \label{Figure:1}%
\end{figure*}

\citet{kuridze2019} used observations of coronal loops filled with material that evaporated as a result of a flare event in order to infer \BLOS\hspace{1pt}values using the weak-field approximation \citep[WFA;][]{landi,Centeno2018}. The spectropolarimetric observations used to carry out the inference were obtained with the  CRisp Imaging SpectroPolarimeter \citep[CRISP;][]{Scharmer_2008} at the Swedish 1-m Solar Telescope \citep[SST;][]{sst} in the \ion{Ca}{II} 8542 \AA\ line. The spatial sampling of the observations was 0.057\arcsec~pixel$^{-1}$, with a cadence of about 32 s. In order to increase the signal-to-noise ratio (S/N), a spatial average was done, resulting in a 0.23\arcsec~pixel$^{-1}$ spatial resolution. We note that \BLOS\hspace{1pt}values were inferred up to heights of 18~Mm above the visible limb, with values at lower heights (up to 9 Mm) peaking at two values, 150 G and 300~G, because of the presence of different structures in the field of view. At upper heights, the inferred \BLOS\hspace{1pt}values were lower, ranging from 50~G to 180~G,  having a median of 90 G.

In a previous paper \citep[hereafter  \citetalias{2020kri}]{2020kri}, we developed a method to infer the \BLOS\hspace{1pt} value using the WFA through a Bayesian scheme, which was applied to off-limb and on-disc spicules in observations obtained in the \ion{Ca}{II} 8542 \AA\ line with CRISP.  We were able to study the \BLOS\hspace{1pt} value both in the vicinity of active regions and in quiet Sun regions. The study in both cases yielded an abundance of magnetic fields of hundreds of Gauss in these structures.

In this paper, we used spectropolarimetric observations (Sect.~\ref{sec:obs}) in the \ion{Ca}{II} 8542 \AA\ line of the low coronal portion of an active region loop bundle in order to infer the total magnetic field strength from the coronal rain clumps that were falling towards the chromosphere. We adapted the method developed in \citetalias{2020kri}, combining the use of the WFA with velocity measurements to infer the magnetic field strength in the lower corona (Sect.~\ref{sec:Data}). Magnetic field strengths of hundreds of Gauss were abundantly inferred (Sect.~\ref{results}), with the possibility of even higher values being present. Concluding remarks are presented in Sect.~\ref{conclusions}.

 \section{Observations} \label{sec:obs}

The observations used in this work were obtained with the CRISP instrument at the SST on  03 June 2016 and consisted of imaging spectropolarimetry in the \ion{Ca}{II} 8542 \AA\ line and in the hydrogen H$\alpha$ line with a pixel scale of 0.057\arcsec~pixel$^{-1}$. The spectral sampling was 0.05~\AA\ in the line centre and 0.25~\AA\ at the wings, with a total of 15 spectral positions for the calcium observations. The spectral sampling of the  H$\alpha$ observations was 0.1~\AA\ at the line centre and 0.25~\AA\  at the wings, totalling 18 spectral positions.  Data reconstruction made use of the Multi-Object Multi-Frame Blind Deconvolution \citep[MOMFBD;][]{VanNoort2005} method. The CRISP data reduction pipeline \citep[CRISPRED;][]{2015delacruz} was used for additional data reduction, including the cross-correlation method of
\citet{Henriques2012A}. The resulting temporal cadence was 36.33s.
  
 The data set analysed consisted of observations obtained between 16:27 UT and 17:27 UT, totalling 100 frames of  spectropolarimetric observations of  off-limb chromospheric structures near active region NOAA AR 12551, as it was approaching the solar west limb. Due to the presence of an active region, coronal rain was easily seen off-limb as small clump-like features falling through the low corona. Given their density and size, the \ion{Ca}{II} 8542 \AA\ intensity signal from these structures can be much lower than the average intensity of the visible disc when looking at an intensity image. Figure~\ref{Figure:1} shows a temporal  frame in the \ion{Ca}{II} 8542 \AA\ line together with a simultaneous image in the H$\alpha$ line. The movie associated with Fig.~\ref{Figure:1} shows the superposed \ion{Ca}{II} 8542 \AA\  and H$\alpha$ intensity over the course of the observations.

One would normally expect that pixels outside the solar disc and off-limb features such as coronal rain, prominences, or spicules would have no signal; however, due to scattering in the Earth's atmosphere and stray light in the telescope, there is a presence of diffuse stray light \citep{1998SoPh..183..229C,Luc2019}. This stray light is averaged over the Sun and it produces a normal spectrum and hence an absorption line. When the level of stray light was averaged over time in each pixel, a declining gradient as a function of the distance from the limb was clearly seen. With this declining gradient, the level of stray light was estimated and removed from the \ion{Ca}{II} 8542 \AA\ intensity.

  \section{Data analysis} \label{sec:Data}
In this section, we describe the different methods used to infer several physical parameters from the observational data.
\subsection{Line-of-sight magnetic field component } \label{sec:bdet}
The main goal of this study was to apply the same method developed in  \citetalias{2020kri} using the WFA to infer the value of  \BLOS. The method is based on the expression that relates the intensity and the Stokes $V$ profile,

\begin{equation}
V (\lambda) = -4.67 \times 10^{-13} \overline{g} f \lambda_{0}^{2} B_{\mathrm{LOS}} \frac{\partial I (\lambda)}{\partial \lambda},
\end{equation}\label{equation:1}

\noindent where $f$ is the magnetic filling factor \citep{landi,AsensioRamos2011},  $\overline{g}$ is the effective Land\'e factor of the line \citep{Landi1982},  $\lambda_0$ is the wavelength of the transition (in angstroms), and $B_{\mathrm{LOS}}$ is given in Gauss. Two conditions need to be satisfied in order to correctly use the WFA: The Zeeman width must be much smaller than the Doppler width, and the magnetic field needs to be uniform along the line of sight (LOS). The first condition is mathematically expressed as

\begin{equation}
 \frac{\overline{g} \Delta \lambda_{Z}}{\Delta \lambda_{D}} \ll 1, \label{eq:ineq}
\end{equation}
\noindent where $\Delta \lambda_{D}$ is the Doppler width of the line and $\Delta \lambda_{Z}$ is the Zeeman splitting.  The Zeeman splitting (in angstroms) is given by the expression 

\begin{equation}
\Delta \lambda_{Z} = 4.67\times 10^{-13} \lambda_0^2 B , \label{eq:zeeman}
\end{equation}

\noindent where $B$ is the field strength (in Gauss). The Doppler width, assuming that the instrument width for CRISP is negligible, is expressed as
\begin{equation}
\Delta \lambda_{D} =\frac{\lambda_0}{c} \sqrt{1.663 \times 10^{-2} \frac{T}{\mu} + \xi^2} , \label{eq:doppler}
\end{equation}

\noindent where $T$ is the temperature (in Kelvin), $\xi$ is the microturbulent velocity (in km $\mathrm{s}^{-1}$), $c$ is the speed of light, and $\mu$ is the atomic weight of the atom from which the spectral line originates. By inserting Eq.~(\ref{eq:zeeman}) into Eq.~(\ref{eq:ineq}), we arrive at the requirement that
\begin{equation}
B \ll \frac{\Delta \lambda_{D}}{ 4.67\times 10^{-13}\overline{g} \lambda_0^2}. \label{eq:maxfield}
\end{equation}

In  \citetalias{2020kri},  it was not possible to infer the value of B. This prevented these authors from checking whether Eq.~(\ref{eq:maxfield}) was satisfied, so they used typical values for $T$ and $\xi$ in order to estimate, with the help of Eq.~(\ref{eq:doppler}), an upper limit for the values of $B$ that could be safely inferred, arriving at the expression

\begin{equation}
B \ll 2650\hspace{1pt}\mathrm{G}. \label{eq:maxfield_kri}
\end{equation}

 In this study, we were able to go a step further because we inferred $B$ and not just \BLOS \hspace{1pt} for some of the coronal rain clumps. By fitting the intensity profile in each inference with a Gaussian model, we approximated the value of the fitted width ($\sigma_{I}$) as the value of the Doppler width. This allowed us to check in every inference, for which $B$ could be obtained, whether Eq.~(\ref{eq:maxfield}) was satisfied. If the inferred value of $B$ was larger than 15 \% the value of the right-hand side of Eq.~(\ref{eq:maxfield}), the result was discarded.

The second necessary condition for the application of the WFA concerns the uniformity of the magnetic field along the LOS. A gradient of the magnetic field along the LOS can produce asymmetries in the shape of the Stokes $V$ profile, and the method developed in \citetalias{2020kri} limits the allowed amplitude asymmetry to only those cases when it could be caused by the noise present in the data. This method was carried out pixel by pixel with a procedure that analysed the correlation between $V$ and $ \partial I (\lambda)/\partial \lambda$ and the amplitude asymmetry of $V$, before proceeding with the inference of $B_{\mathrm{LOS}}$. A Bayesian inference scheme was developed in order to obtain a posterior probability distribution of $B_{\mathrm{LOS}}$, from which the mode and the 95\% highest posterior density (HPD) interval were extracted as statistical results.
\begin{figure}
   \centering
   \includegraphics[width=8cm]{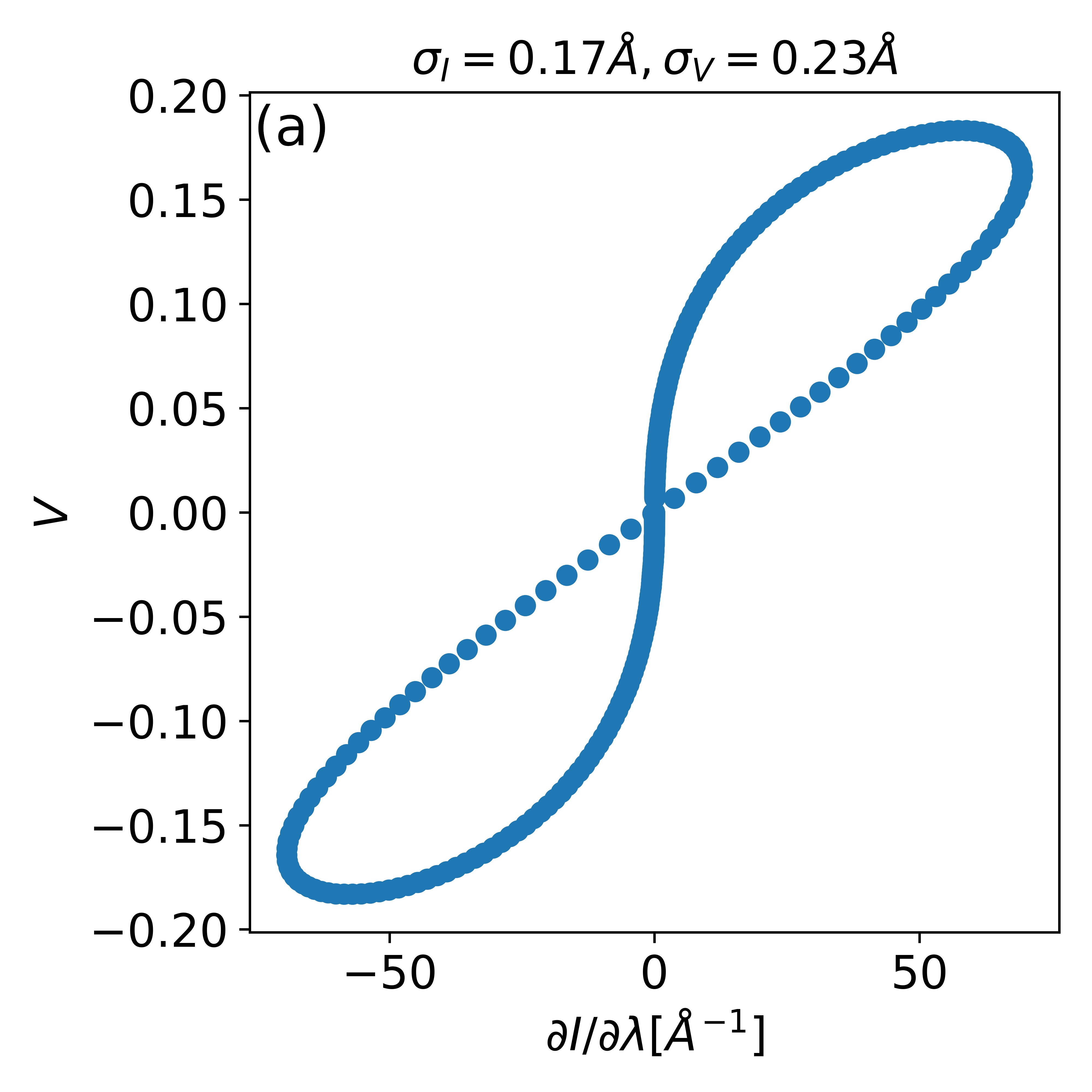}
   \includegraphics[width=8cm]{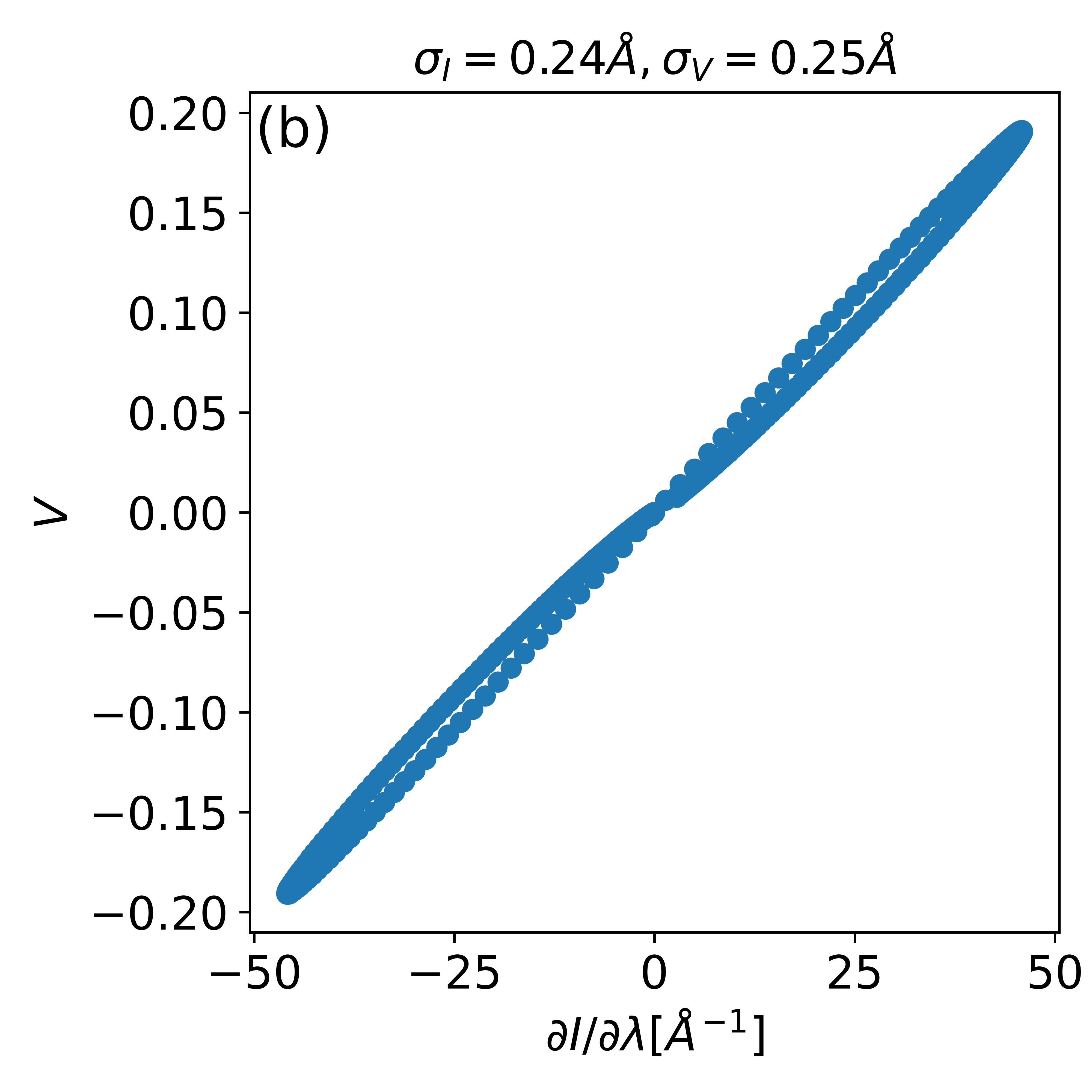}

   \caption{(a) Plot of $V$ as a function of $\partial I /\partial \lambda$ displaying the shape of an '8' due to the discrepancy between $\sigma_V$ and $\sigma_I$, whose values are given on top of the frame. (b) Same as (a), but for a case in which the most significant digit of $\sigma_V$ and $\sigma_I$ is the same.}
              \label{Figure:2}%
\end{figure}
 \begin{figure*}
   \centering

\includegraphics[width=18cm]{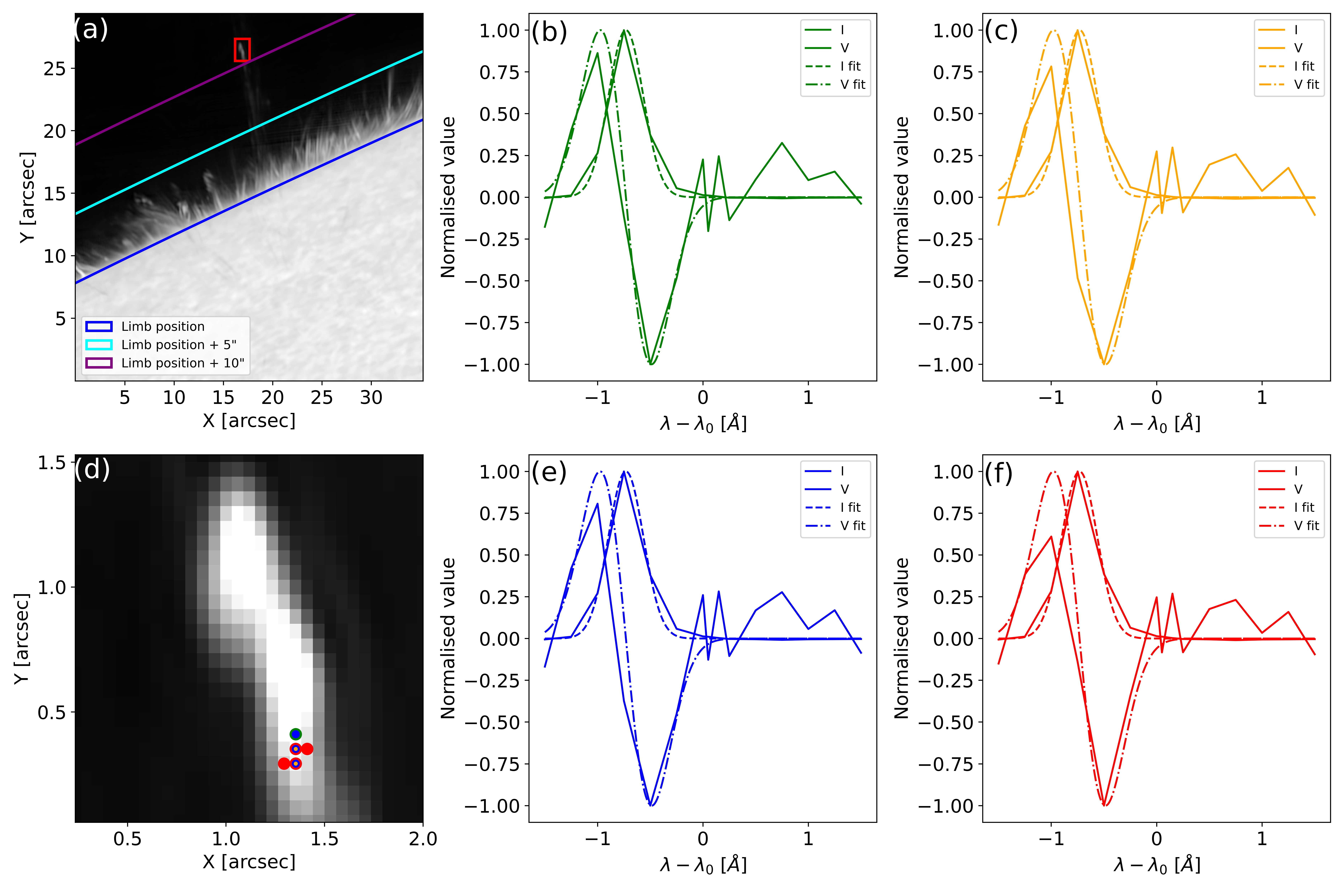}
   \caption{Example of a coronal rain clump present in the observations and the averaging of the Stokes $I$ and $V$ profiles in groups of adjacent pixels. (a) Coronal rain clump at $-0.75 $ \AA \hspace{1pt}from the line centre from an image taken at  16:40:55 UT. (d) Magnified view of the red rectangle of panel (a) in order to enhance the view of the clump. (b), (c), (e), and (f) Stokes $I$ and $V$ profiles normalised to their respective maxima as they appear when averaging the signal from one, two, three, and four pixels, respectively. The colour of the $I$ profile matches the colour of the pixel (or group of pixels) on which the signal is averaged in order to generate the profile as shown on (d). The continuous lines reflect the observed profiles, while the dashed lines depict the profiles that result from fitting $I$ and $V$ with a Gaussian profile and the derivative of a Gaussian profile,  respectively.}
              \label{Figure:3}%
\end{figure*}
 In \citetalias{2020kri}, this analysis was applied to spicules and was carried out for each pixel; however, given the weakness of the Stokes $I$ and $V$ signals present in coronal rain pixels and the rigorous inference scheme developed, a pixel-by-pixel study applied to coronal rain yielded no pixels satisfying the criteria imposed. A modification of the method devised by these authors was developed in order to circumvent this issue. Both the Stokes $I$ and $V$ profiles of each pixel (or set of pixels, see next paragraph) inside a clump were fitted with model functions. In the case of Stokes $I$, the model profile was a Gaussian function, while the Stokes $V$ profiles were fitted with the derivative of a Gaussian profile. The Stokes $I$ fit had three free parameters: the amplitude ($a_{I}$), the width ($\sigma_I$), and the position of the peak ($\lambda_I$) of the Gaussian profile. The Stokes $V$ fit had two free parameters: the amplitude ($a_{V}$) and the width ($\sigma_V$) of the Gaussian profile. The position of the peak of the $V$ fit  was held fixed at the position that resulted from fitting the $I$ profile ($\lambda_V= \lambda_I$). Two criteria were imposed on the results of both fits.  The first criterion was to set an upper limit of $10^{-3}$ for the sum of the difference between the value of the fitted $V$ profile and the observed $V$ profile. The second criterion was based on the similarity between the resulting widths of both fits. When they were too dissimilar, the relation between $V$ and $\partial I$ /$\partial \lambda$ was seen to take an '8' shape (see Fig.~\ref{Figure:2}). We tried to avoid this as much as possible by limiting the analysis to those profiles for which the most significant digit of both widths was identical. After both criteria were satisfied, the same asymmetry criterion imposed in \citetalias{2020kri} was imposed on the observed $V$ profile. If the asymmetry criterion was also satisfied, the value of $B_{\mathrm{LOS}} $ was inferred.

This analysis was first carried out pixel by pixel, but given the small number of results,  a spatial averaging of neighbouring pixels was done. For each clump, the set of pixels inside it was determined, and all the possible groups of $n$ connected pixels (with $n$= 1, 2, 3, and 4) was used to obtain average Stokes $I$ and $V$ profiles. These profiles were then subjected to the analysis described above. Figure~\ref{Figure:3} shows an example of the type of profiles that resulted when performing this analysis for all the values of $n$. The normalisation of the profiles shown in  Fig.~\ref{Figure:3} depends on their maximum value, which is different for the various values of $n$. As a result, it is not visually obvious that this averaging leads to an increase in the S/N as $n$ is increased. Nevertheless, in general, performing a spatial average increased the ratio of the extrema of the $V$ lobes to the extrema of the noise that was present at wavelengths located far from the intensity peak.

 A single pixel could pertain to a group of any $n$ pixels at the same time. Furthermore, groups with common pixels, but with a different value of $n$ were considered different groups. This was done to maximise the different possible outcomes and to avoid the need to single out the best set of groups of $n$ pixels that had no overlapping pixels between groups, which is something that could prove difficult to realise with an automatic criterion.

\subsection{Velocity measurement} \label{Sect:v}

The line-of-sight velocity component ($v_{\mathrm{LOS}}$) was estimated from the Doppler shift of the intensity profile  for each clump. A mean intensity profile was obtained by averaging over the intensity signal of the pixels in the clump, and a Gaussian fit was performed in order to estimate the position of the intensity peak. The error in this parameter value was estimated from the covariance matrix of the fit, and it was then propagated to the error in $v_{\mathrm{LOS}}$.

If a coronal rain clump could be spotted unequivocally at least twice, the plane-of-the-sky component of the velocity ($v_{\mathrm{POS}}$) was estimated by measuring its displacement on the field of view. For this purpose, the centre of mass of the clump was estimated by weighting each pixel according to its peak intensity:

\begin{equation} 
X_{\mathrm{CM}} = \frac{\sum_{k=0}^{N} x_k i_k}{\sum_{k=0}^{N} i_{k}}, \hspace{1cm} 
Y_{\mathrm{CM}} = \frac{\sum_{k=0}^{N} y_k i_k}{\sum_{k=0}^{N} i_{k}},
\end{equation}

\noindent where $N$ is the total number of pixels in the clump, $x_k$ and $y_k$ are the coordinates of each pixel on the plane of the sky, $i_k$ is the peak intensity of each pixel, and $X_{\mathrm{CM}}$ and $Y_{\mathrm{CM}}$ are the clump centre-of-mass coordinates. Assuming a straight trajectory, the displacement of the centre of mass divided by the elapsed time gives an approximation of $v_{\mathrm{POS}}$. The error in this measurement  was propagated from the error in $i_k$, which in turn comes from the Gaussian fit and by approximating the uncertainty of the $x$- and $y$-coordinates of each pixel to be one pixel. 

\subsection{Plane-of-the-sky magnetic field component}\label{Sect:bpos}

Assuming that the plasma moves along magnetic field lines, after obtaining $v_{\mathrm{LOS}}$, $v_{\mathrm{POS}}$, and \BLOS,\hspace{1pt}we estimated the plane-of-the-sky magnetic field component ($B_{\mathrm{POS}}$) with the following relation:

\begin{equation}
B_{\mathrm{POS}} = B_{\mathrm{LOS}} \frac{v_{\mathrm{POS}}}{v_{\mathrm{LOS}}}.
\end{equation} \label{eq:3}

\subsection{Magnetic field strength}

 Once the plane-of-the-sky magnetic field component was obtained, the magnetic field strength was inferred as follows:

\begin{equation}
B= \sqrt{B_{\mathrm{LOS}}^2+ B_{\mathrm{POS}}^2} = B_{\mathrm{LOS}} \times \sqrt{1+ \frac{v^{2}_{\mathrm{POS}}}{v^{2}_{\mathrm{LOS}}}}. \label{eq:4}
\end{equation}

 \noindent Equation~(\ref{eq:4}) was introduced in the Bayesian scheme developed in \citetalias{2020kri}  in order to obtain a posterior distribution of $B$ and its 95\% HPD interval. As is explained in Sect.~\ref{Sect:bpos}, the errors in \vPOS\hspace{1pt}and \vLOS\hspace{1pt}were propagated into the value of their squared ratio as it appears in Eq.~(\ref{eq:4}), which was then introduced as a new parameter into the Bayesian scheme with a uniform prior bounded by its minimum and maximum values according to its uncertainty.

\section{Results and discussion} \label{results}

In this section, we present the results of the analysis of the different physical parameters described in Sect.~\ref{sec:Data}.
\begin{figure}
   \centering
   \includegraphics[width=7cm]{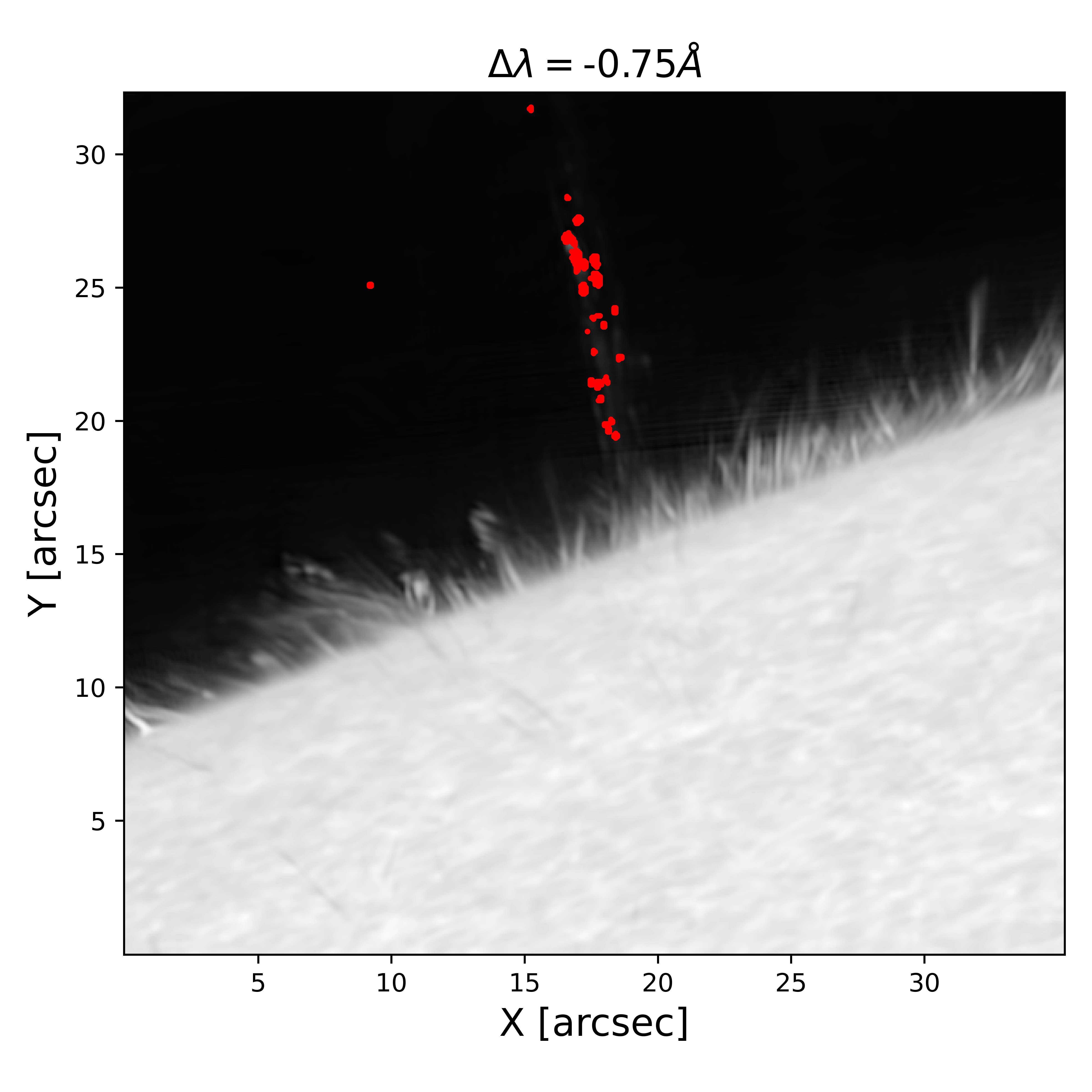}

   \caption{ Logarithm of the intensity at 16:40:55 UT in the same way as Fig.~\ref{Figure:1}. The red dots correspond to the positions where the magnetic field was inferred and/or a clump was detected in at least one scan.}
              \label{Figure:4}%
\end{figure}
\subsection{Behaviour and shape}

Throughout the whole duration of the observations, which added up to an hour, clumps of multiple shapes were seen to mainly fall in a localised region of the field of view, following a very similar path (see movie accompanying Fig.~\ref{Figure:1}). Their most common shape displayed a bright centre with fading tails both in front or behind them along their trajectory. Sometimes they displayed nearly circular shapes, and it was also common to observe the complete lack of a bright centre in favour of a much longer thin structure that covered the whole path usually followed by the clumps. Several clumps also varied in shape along their trajectory, becoming more or less elongated as they fell into the chromosphere, which is a phenomenon that was also observed in H${\mathrm{\alpha}}$ spectroscopic observations by \citet{2012ApJ...745..152A}.

Although the actual number of visible clumps was much larger, we only analysed and identified those that were mostly isolated, omitting  those that were packed too closely together to discern a visible boundary between them. A total of 37 clumps were identified, mostly along a common path with a few exceptions, as Fig.~\ref{Figure:4} shows.

The appearance of clumps in the \ion{Ca}{II} 8542 \AA\ line and in the H$\alpha$ line is very different. As Fig.~\ref{Figure:1} shows,  when seen in H$\alpha$, the rain has a stronger signal compared to the disc intensity. The rain is also much more abundant in this spectral line than in the \ion{Ca}{II} and in several cases an elongated clump is seen in H$\alpha$, while only its dense core is seen in the corresponding \ion{Ca}{II} observation. This disparity arrives at the extreme case of some clumps clearly spottable in $\mathrm{H\alpha}$, while not being visible in \ion{Ca}{II}.
\begin{figure}
\centering
\includegraphics[width=8cm]{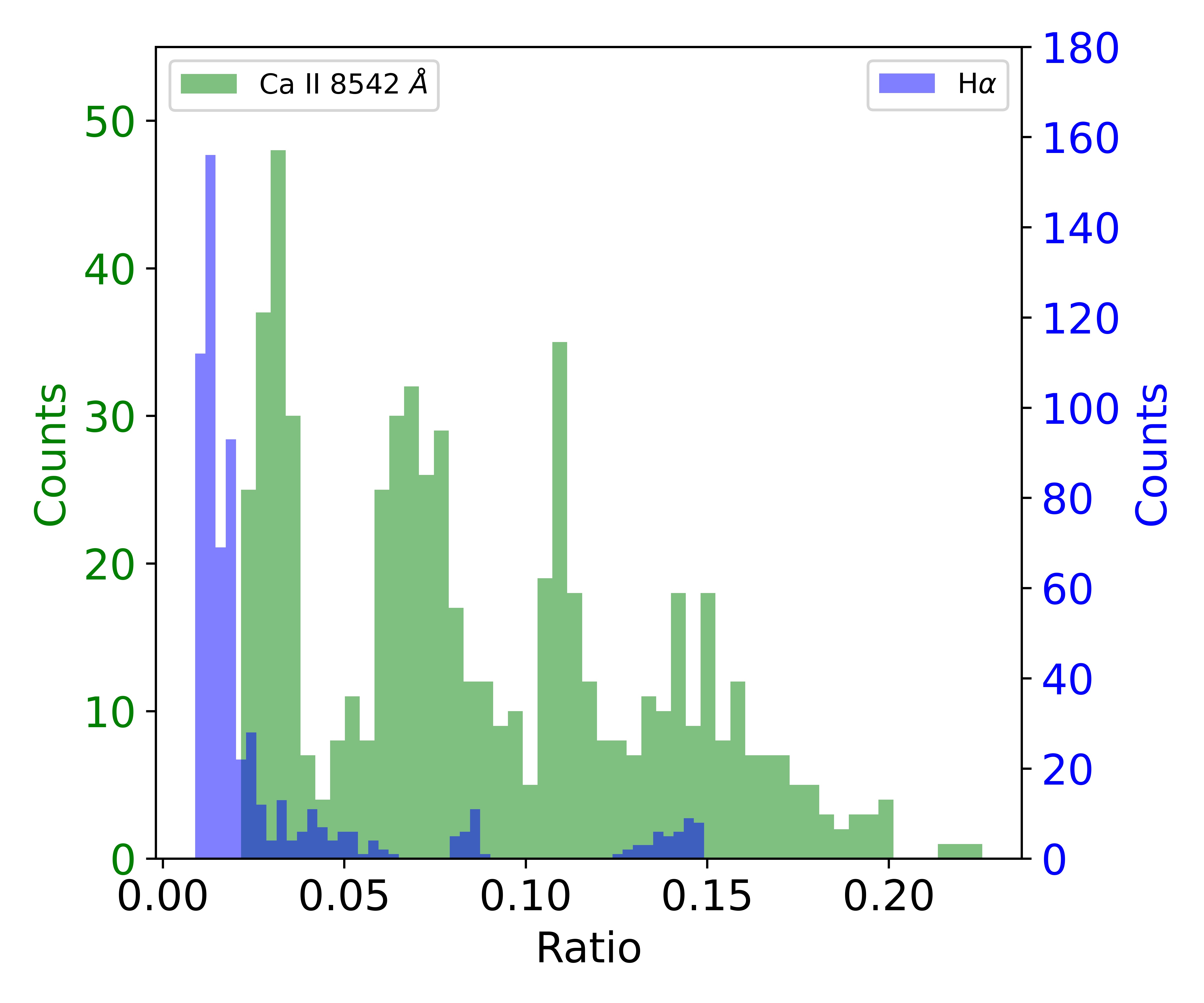}
 \caption{Ratio of the average stray light profile to the $I$ profile at the position of the intensity peak for the pixels used in the magnetic field inference. The  \ion{Ca}{II} 8542 \AA\ results are shown in green and the H$\alpha$ results are shown in blue.  }
              \label{Figure:5}%
\end{figure}
\subsection{Stray light}

In order to quantify the importance of the removal of the average stray light profile from the observations, we calculated the ratio of this average profile to the $I$ profile at the position of the intensity peak for the \ion{Ca}{II} 8542 \AA\ and H$\alpha$ observations for the pixels in which coronal rain could be  simultaneously detected in these two spectral lines. The results are shown in Fig.~\ref{Figure:5}. The stray light profile amounted up to 23\% of the observed profile for the \ion{Ca}{II} 8542 \AA,\, while it only amounted up to 15\% for H$\alpha$, in most cases being less than 5\%. Therefore, stray light correction appears to be a more important effect when studying coronal rain in the \ion{Ca}{II} 8542 \AA\ line.
\begin{figure}
   \centering
   \includegraphics[width=8cm]{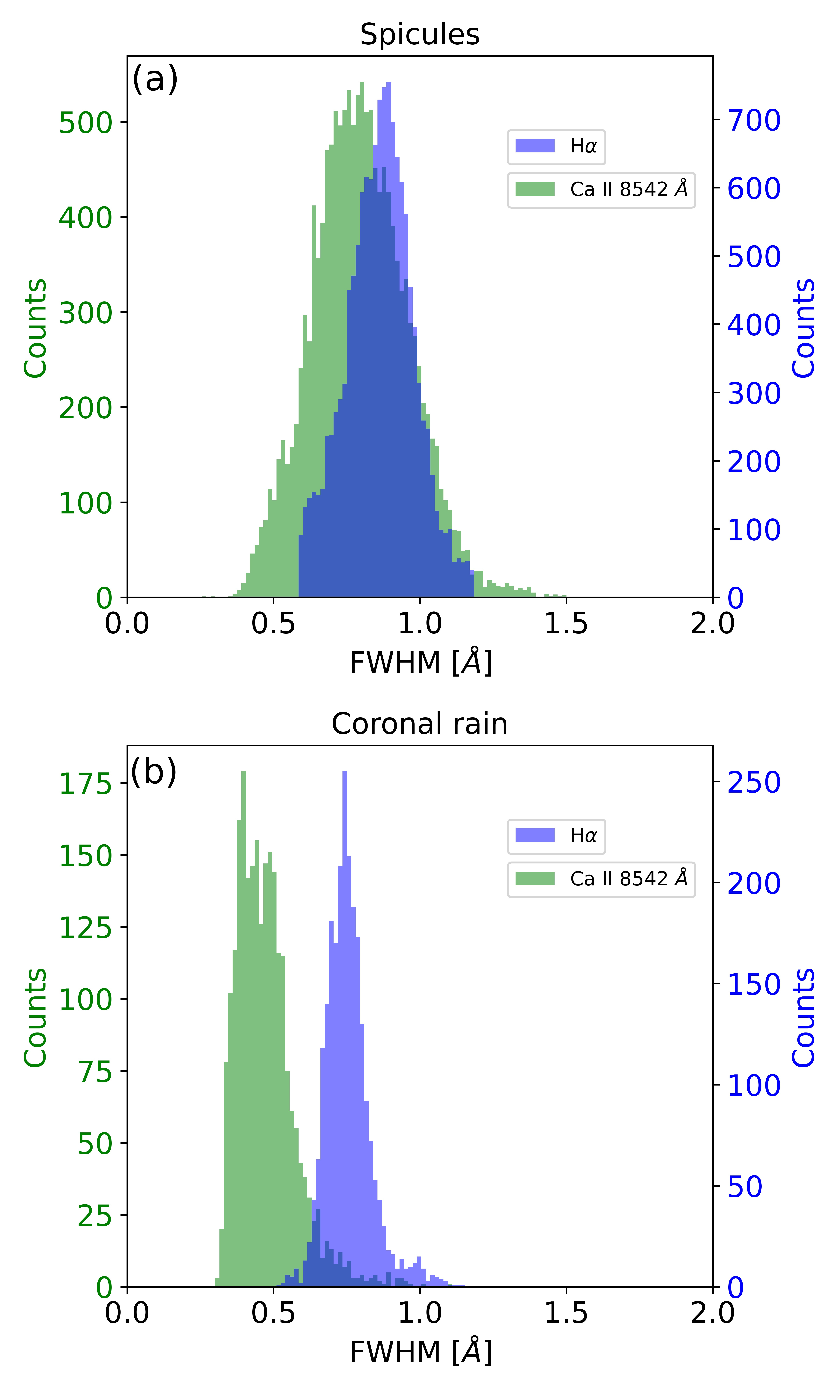}

   \caption{Distributions of the measured values of the Stokes~$I$ FWHM of the \ion{Ca}{II} 8542 $\AA$ (in green) and H$\alpha$ (in blue) lines, using a Gaussian fit for (a) spicules and (b) coronal rain clumps.}
              \label{Figure:6}%
\end{figure}

\subsection{Intensity width} \label{sect:43}

In order to study the average width of the coronal rain intensity profiles, we obtained the full width at half maximum (FWHM) by performing a Gaussian fit to the \ion{Ca}{II} 8542 $\AA$ Stokes $I$. Additionally, we performed Gaussian fits to the intensity profiles of the spicules present in the data set analysed here\footnote{The data set of this paper is \#8 of \citetalias{2020kri}.}. The results are shown in Fig.~\ref{Figure:6}. We found that the ratio of the average width of the intensity profile of spicules to that of the intensity profile of the coronal rain clumps for the \ion{Ca}{II}~8542~$\AA$ line was 1.7, while its value for the H$\alpha$ line was 1.1. The difference in ratios between the \ion{Ca}{II} 8542 $\AA$ and H$\alpha$ lines could be caused by the fact that, for a given temperature, the thermal component of the Doppler width is a factor of 40 larger for  H$\alpha$ because of the difference in atomic mass between hydrogen and calcium atoms.

\begin{figure}
   \centering
   \includegraphics[width=8cm]{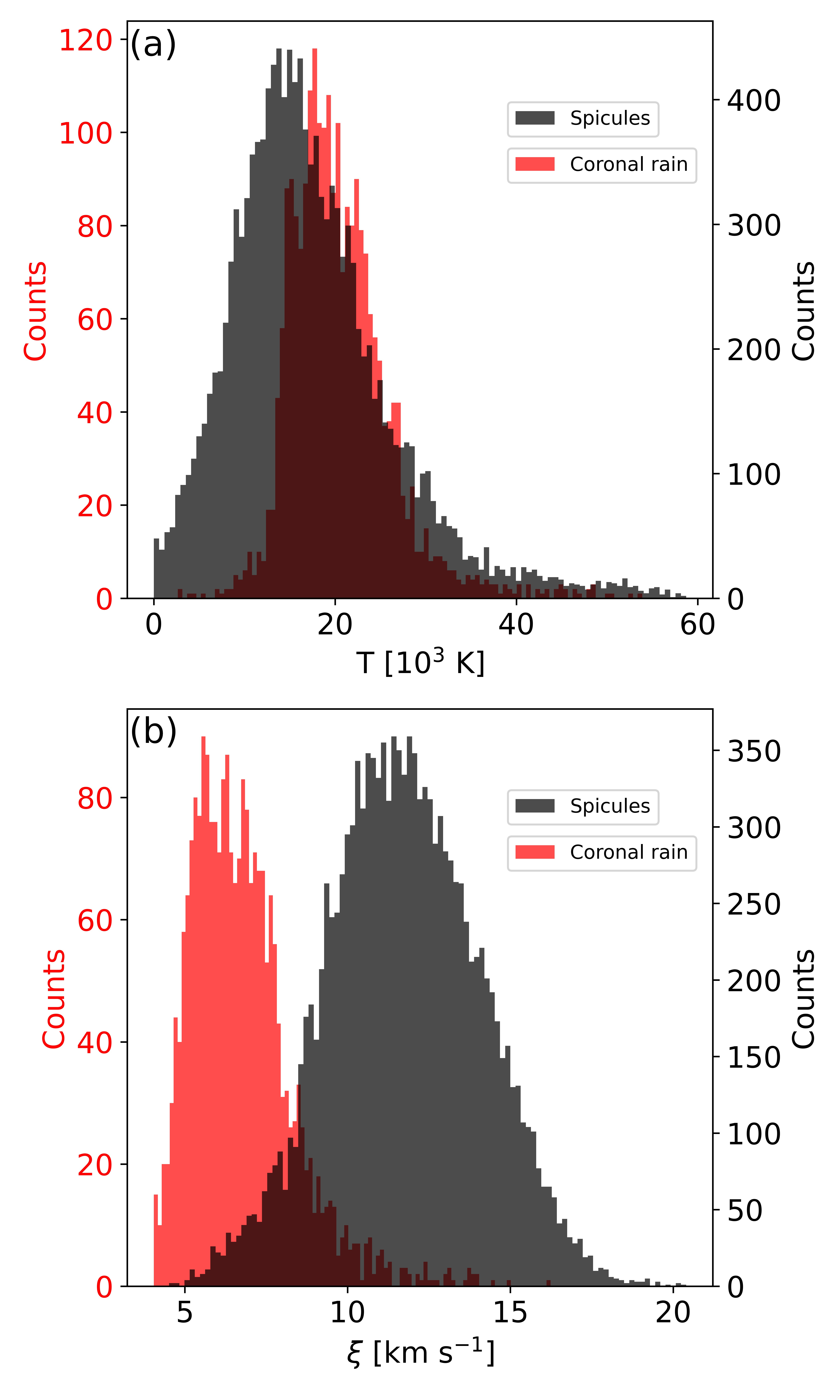}

   \caption{Distributions of the measured values of the (a) temperature and (b) microturbulent velocity of coronal rain clumps (in red) and spicules (in black) using Eq.~ (\ref{eq:doppler}).}
              \label{Figure:7}%
\end{figure}
\begin{figure*}
   \centering
   \includegraphics[width=6cm]{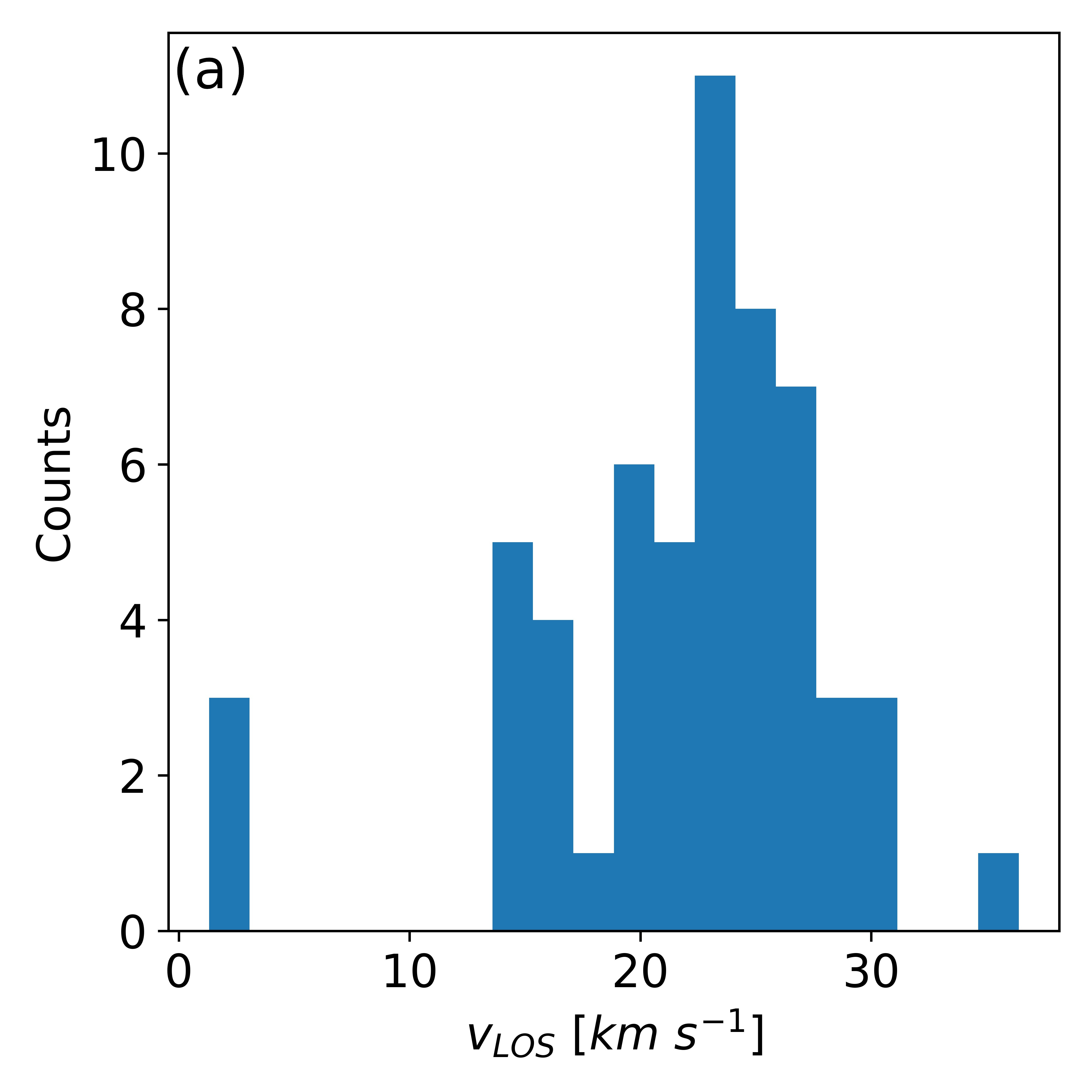}
   \includegraphics[width=6cm]{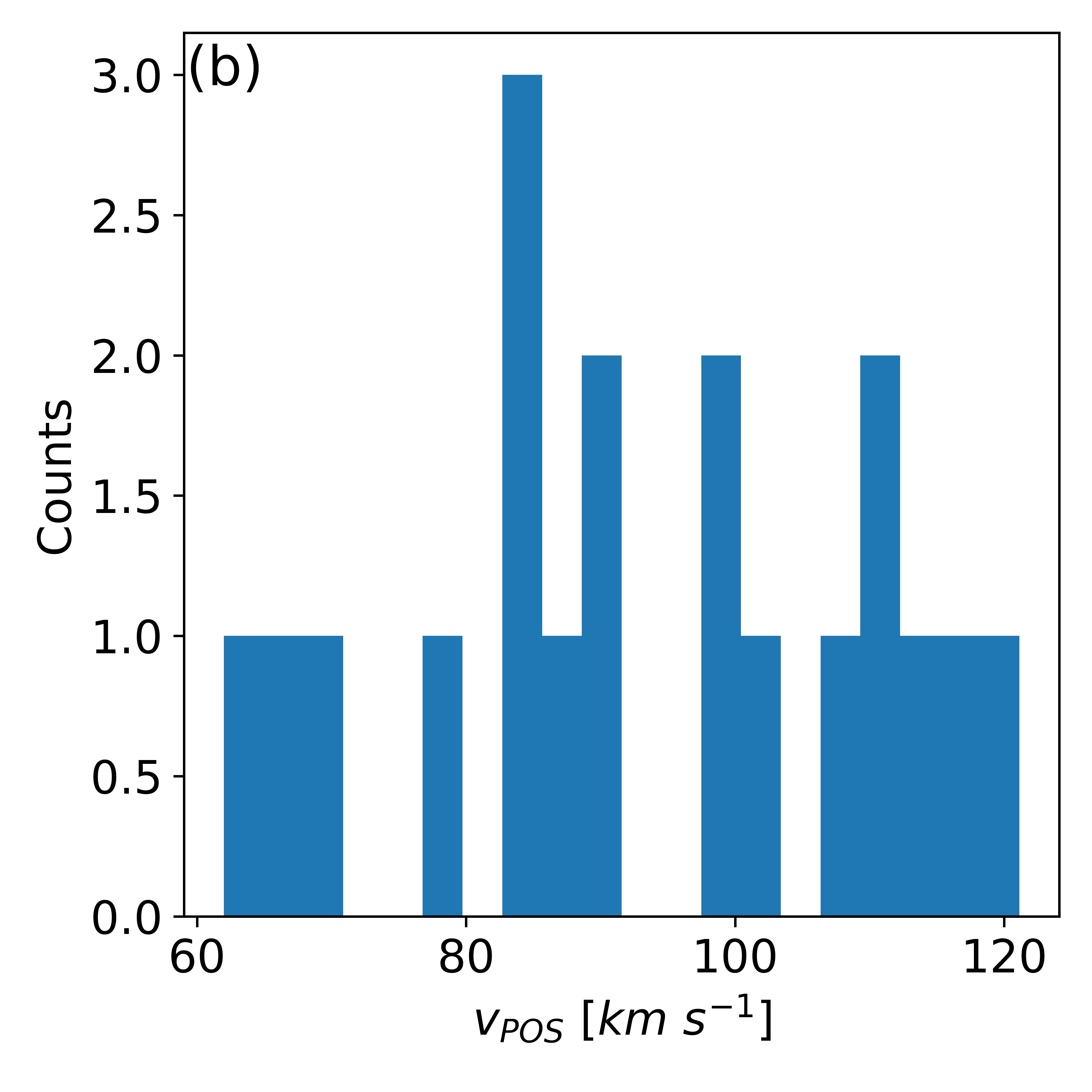}
   \includegraphics[width=6cm]{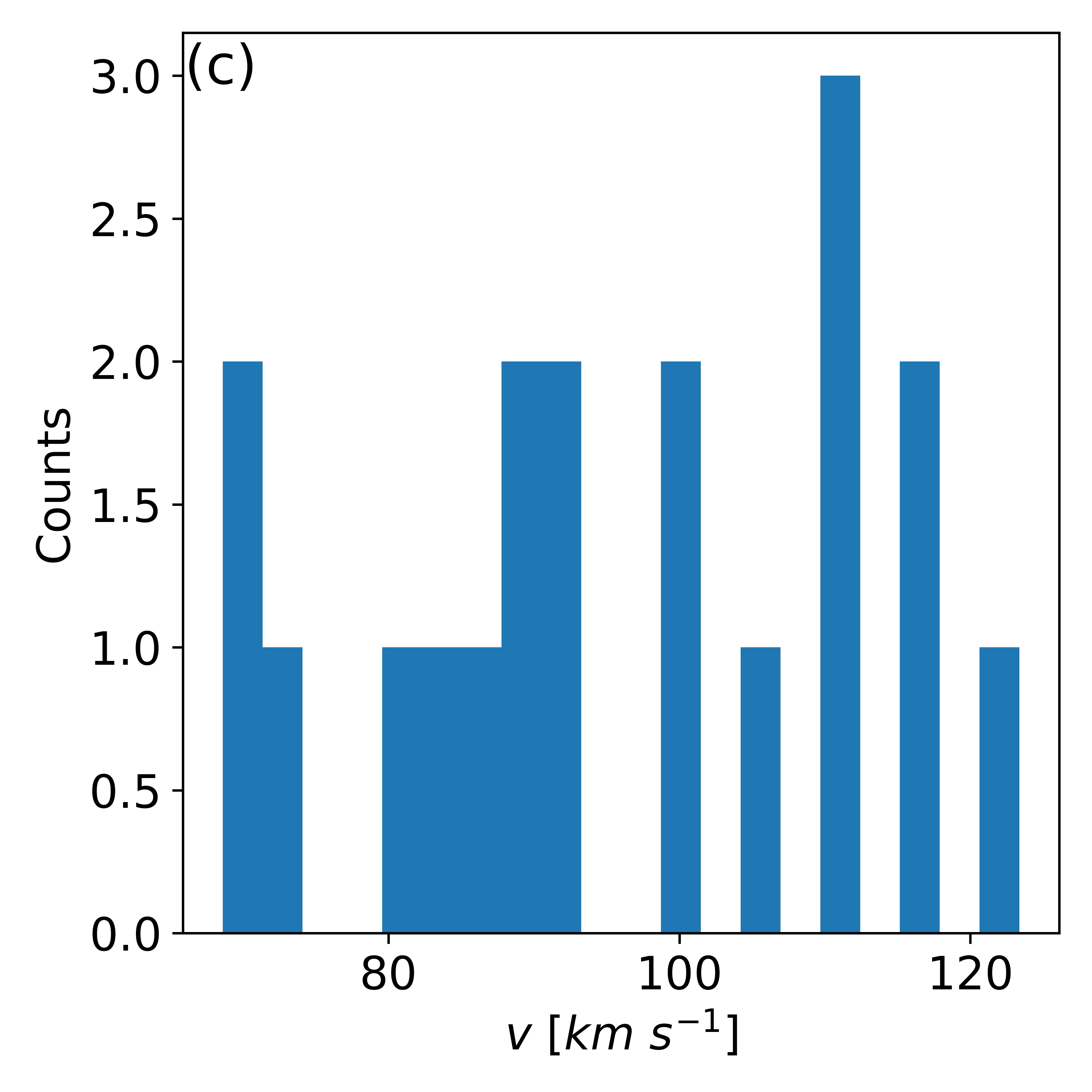}
   \caption{ Occurrence of the different values of (a) \vLOS \hspace{1pt}and (b) \vPOS \hspace{1pt}as inferred from the Gaussian fit of the intensity and (c) inferred values of the total velocity calculated with Eq.~(\ref{eq:4}).}
              \label{Figure:8}%
\end{figure*}

Making use of Eq.~(\ref{eq:doppler}), we employed a similar method as the one described by \citet{2020A&A...633A..11F} to measure the temperature and microturbulent velocity of coronal rain clumps using the combined observations in the \ion{Ca}{II} 8542 \AA\ and $\mathrm{H\alpha}$ lines, assuming equal temperatures and microturbulent velocities for both lines. The results are shown in Fig.~\ref{Figure:7}. The measured temperatures were of the order of tens of thousands of kelvin, with an average value of $20\,500$~K, while the microturbulent velocity ranged from 1 to 13 km $\mathrm{s}^{-1}$, with an average value of 6 km $\mathrm{s}^{-1}$.

The method of \citet{2020A&A...633A..11F} to measure $T$ and $\xi$ was also used for the spicules present in the data set. The temperature range was broader than that of coronal rain clumps, with an average value of $19\,900$~K, while the microturbulent velocity had an average value of 13 km $\mathrm{s}^{-1}$, approximately two times the value measured for coronal rain. This difference in non-thermal effects could explain the disparity between the measured values of the FWHM between spicules and coronal rain in the \ion{Ca}{II} 8542 $\AA$ line described before.

\begin{figure}
   \centering

 \includegraphics[width=7cm]{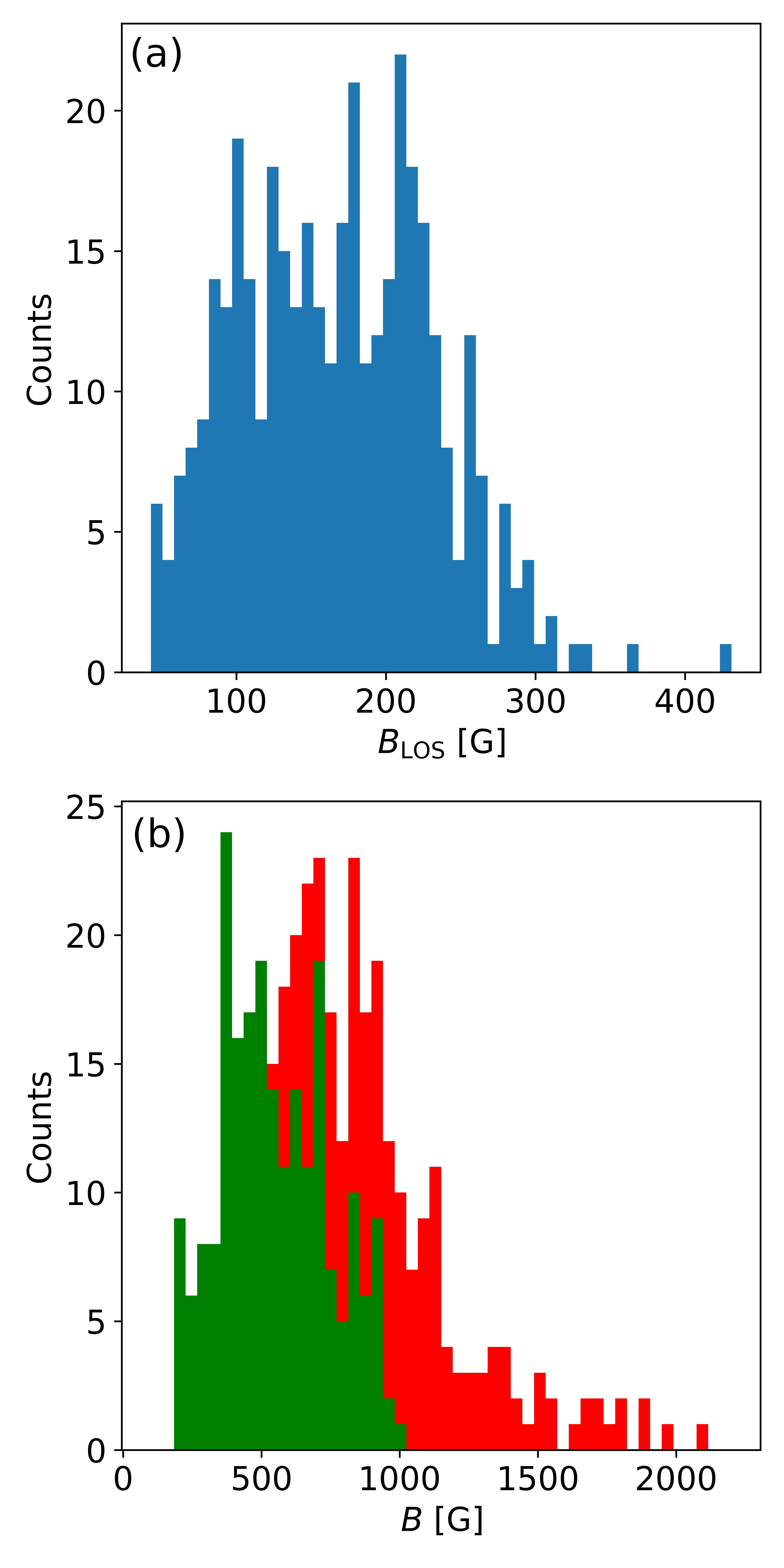}
   \caption{(a)  Inferred \BLOS \hspace{1pt} values using the method described in Sect.~\ref{sec:bdet}. (b) Inferred values of the total magnetic field for the cases in which \vPOS \hspace{1pt} could also be obtained. The colours of panel (b) indicate whether Eq.~(\ref{eq:maxfield}) was satisfied (green) or not (red). }
              \label{Figure:9}%
\end{figure}
\subsection{Velocities}

The identified clumps provided a total of 57 \vLOS  \hspace{1pt}measurements using the method described in Sect.~ \ref{Sect:v}. Given the necessity for the clumps to be spotted at two different times and the limited off-limb field of view, a  \vPOS \hspace{1pt}measurement could be determined for only 19 of these clumps. Another reason for the limited number of \vPOS  \hspace{1pt}measurements was the cadence of the data, which resulted in a large motion of the clumps between two consecutive temporal frames. The results are displayed in Fig.~\ref{Figure:8}. The values of \vLOS\hspace{1pt} ranged from a few km $\mathrm{s}^{-1}$ up to about 35 km~$\mathrm{s}^{-1}$, peaking at around 25 km $\mathrm{s}^{-1}$. The total velocity values were calculated with the expression

\begin{equation}
v = \sqrt{v^{2}_{\mathrm{LOS}} + v^{2}_{\mathrm{POS}}},
\end{equation}

\noindent and they ranged between 70 and 120  km~$\mathrm{s}^{-1}$, with a mean value of 95  km $\mathrm{s}^{-1}$. These values are in excellent agreement with previous studies \citep{2012ApJ...745..152A,2001SoPh..198..325S,2005A&A...443..319D}. The fact that the total velocity achieved values that were much larger than the inferred  \vLOS \hspace{1pt} values indicates that the trajectory followed by most of the clumps was almost parallel to the plane of the sky with a small component in the direction of the observer. Hence, \BLOS\hspace{1pt} values much smaller than $B$ values were expected.

\subsection{Magnetic field inference}

Using the method detailed in Sect.~\ref{sec:Data}, we performed the inference of \BLOS \hspace{1pt} for different values of $n$ (1, 2, 3, and 4). We also performed the inference of $B$ for those clumps that had a \vPOS \hspace{1pt}  value assigned. As Fig.~\ref{Figure:4} shows, most of the observed clumps were falling along a common path. This fact, combined with the observation that the measured ratio of \vLOS \hspace{1pt}values to the  \vPOS \hspace{1pt}values was located mostly between 0.2 and 0.3, with an average of 0.23, allowed us to use this average value as a representative value for the clumps falling along this common path, but that are only seen in one scan. The results are shown in Fig.~\ref{Figure:9}. The inferred \BLOS \hspace{1pt}values were found to be between 40 and 430 G, mostly between 100 and 220 G. The distribution of $B$ ranged from 170 up to 980~G. The average inferred $B$ value was 550 G. As evidenced by Fig.~\ref{Figure:9}, $B$ values as large as 2085 G were originally inferred. However, many of the results, especially those corresponding to large values of $B$, had to be discarded in order to satisfy Eq.~(\ref{eq:maxfield}).

The distributions displayed in Fig.~\ref{Figure:9} combine the total results for all the different values of $n$. Table~\ref{table:Table1} summarises the results separately for each $n$ and for all $n$ combined. Except for the average inferred value of $B$ for $n=2$, it seems that the inferred values of both \BLOS \hspace{1pt} and $B$ decrease as the value of $n$ increases. One of the possibilities could be the superposition effects that can take place when combining signals from different pixels, which is something that was explored in detail in \citetalias{2020kri} when applying this analysis on signals originating from spicules. These authors concluded, albeit for spicular material, that when spatial or temporal averages are undertaken, the inferred values of the magnetic field strength tend to decrease.

Comparing the values of \BLOS \hspace{1pt}  inferred using our inference method with the results presented by \citet{2016ApJ...833....5S}  or \citet{kuridze2019} is challenging because they are heavily influenced by the orientation of the loop with respect to the observer and, therefore, they can vary strongly depending on the observational data used. We do, however, infer similar values to those reported by \citet{kuridze2019} using a different method to ours but also employing the WFA, which is something that indicates that both studies report the abundance of magnetic field values of hundreds of Gauss.
\begin{table}
         \renewcommand*{\arraystretch}{1.1}
        \centering
        \caption{ Values (in Gauss) of the average \BLOS, $B$, and their corresponding 95\% HPD intervals for the different values of $n$.}
        \label{table:Table1}
        \begin{tabular}{ccccc}
        
            \hline \hline
             n & \BLOS & HPD(\BLOS) & $B$&HPD($B$)\\
            \hline
            1&210&[201, 215]&585& [543, 636]\\

            2&190&[178,196]&507&[490, 545]\\

            3&174&[164, 181]&564&[538, 611]\\

            4&156&[149, 161]&542&[518, 586]\\

            1, 2, 3, 4&167& [158, 173]&548& [523, 592]\\\hline
        \end{tabular}
\end{table}
As mentioned in the Introduction, \citet{2016ApJ...833....5S} inferred the values of $B$ in two different ways: from the inferred \BLOS \hspace{1pt} values using the $\mathrm{HELIX}^{+}$ code combined with the triangulated geometry of the loop and using the Hanle-slab part of the $\mathrm{HELIX}^{+}$ code. Unfortunately, these authors report that the values of $B$ inferred using the second method are not reliable for heights above the visible surface lower than 14 Mm, so it would be unwise to compare our inference results with those obtained through that method because we were only capable of detecting clumps up to 9 Mm above the limb. However, it is possible to compare our results with those obtained using the geometry of the loop combined with the \BLOS \hspace{1pt}inference. Inspecting Fig.~8 of  \citet{2016ApJ...833....5S} while considering the large error in their reported height measurements, we can estimate that the values inferred by those authors at the heights corresponding to those of our observations (heights lower than 9 Mm) are larger than 1000 G. The highest $B$ value that we inferred with our method was around 1000 G, with most values centred around 500 G.  Nevertheless, we were limited to inferring magnetic field values that are compatible with the conditions for the WFA to be applicable. This means that even though magnetic fields as large as those reported by \citet{2016ApJ...833....5S} can very well be present in coronal rain clumps, they cannot be safely inferred using the WFA and the \ion{Ca}{II} 8542 \AA\ line. The distribution of values of $B$ in Fig.~\ref{Figure:9} could lead us to believe that values higher than 1000 G are present in the coronal rain clumps studied in this work.

\section{Conclusions} \label{conclusions}

We have used spectropolarimetric observations in the \ion{Ca}{II}~8542~$\AA$ line with the CRISP instrument installed at the SST in order to study the magnetic field of the low coronal portion of catastrophically cooled active region coronal loops by using coronal rain clumps as magnetic tracers. We have employed the WFA as a simple and computationally inexpensive tool to paint a picture of the type of magnetic fields that are found in this part of the solar atmosphere. However, the WFA can only be applied when certain conditions about the plasma and the magnetic field are satisfied. This means that we needed to make sure that the conditions for its applicability were carefully assessed. The method employed here was a variation in the method developed by \citet{2020kri}, adding a few extra steps related to the goodness of the fit that was performed to the observed Stokes $I$ and $V$ profiles and using velocity measurements to infer the magnetic field strength.

Given their density and size, it is a challenging task to obtain spectropolarimetric signals in coronal rain clumps with the observational capabilities of current instruments. We needed to overcome this issue in order to obtain profitable results by performing spatial averages of multiple connected pixels. We constructed groups of up to four pixels, which were generated randomly inside each clump. By performing these averages, we were able to increase the S/N of the data, increasing the possibility of obtaining magnetic information from coronal rain clumps. This exercise proves that although it is a challenging task, it is possible to obtain spectropolarimetric signals in these structures with the available instrumentation. We expect that with the arrival of telescopes such as DKIST we will be able to gather even more information with better resolution and S/N and paint a more complete picture of the dynamics of coronal rain clumps.

One of the biggest limitations we have faced, besides the presence of noise in the data, was the cadence. With a small off-limb field of view and structures that follow a path mostly parallel to the plane of the sky with velocities of ~100  km $\mathrm{s}^{-1}$, a cadence of~36.33 s is far from ideal. It was difficult to spot a significant number of clumps more than once in time and space in order to approximate their magnetic field strength with the aid of the velocity vector, so the results were limited. Additionally, with a better cadence, we would have been able to obtain more precise information about the geometric structure of the multiple coronal loops that were present in the data, and that would have allowed for a precise study of the variation in the magnetic field strength with height. 

We were also limited by the range of applicability of the WFA regarding the values of the magnetic field strengths that we could be safely inferred. As its name suggests, it can only be applied on weak-field regimes and, beyond a certain limit, we cannot fully trust the values that were inferred. This circumstance hindered our ability to compare our results with those of other authors. It seems clear that a complementary inference method needs to be used parallel to the WFA in order to obtain results regarding the larger magnetic field values suggested by the right tail of the distribution of Fig.~\ref{Figure:9}b. What we have shown is that the WFA is useful to determine the lower to medium range of magnetic field values that seem to be present near the feet of catastrophically cooled coronal loops.  Since the magnetic field strength is expected to decrease with height, the WFA should also be applicable above the 9 Mm limit of our data.

Regardless of the limitations of the method and the observational challenges that we have faced, it seems clear that there is an abundant presence of magnetic fields of hundreds of Gauss in the lower coronal part of catastrophically cooled active region coronal loops, and coronal rain clumps can indeed serve as tools not only to map out the geometry of the loops with their motion, but also as sources of information about the actual magnetic field values in this part of the solar corona.

As shown in Sect.~\ref{sect:43}, we have estimated the temperature and microturbulent velocity of coronal rain clumps and spicules using the method described  by \citet{2020A&A...633A..11F}. Using this method entailed the assumption that the temperature and microturbulent velocity of the plasma seen in the \ion{Ca}{II}~8542~$\AA$ and H$\alpha$ lines are the same. This is an approximation that must be considered when addressing the results obtained. The average temperature of coronal rain clumps and spicules was found to be very similar, albeit the temperature distribution of spicules was much broader. There was a noticeable difference in the measured microturbulent velocity, with the average value for spicules being about two times larger than that of coronal rain clumps. Considering the assumptions made in order to obtain these results, the next step would be to compare them to the results obtained with different approaches, such as the use of non-local thermodynamic equilibrium inversion codes.

\begin{acknowledgements}
MK and RO acknowledge support from the Spanish Ministry of Economy and Competitiveness (MINECO) and FEDER funds through project AYA2017-85465-P. They are also grateful for the travel support received from the International Space Science Institute (Bern, Switzerland) as well as for discussions with members of the ISSI team on ``Observed multi-scale variability of coronal loops as a probe of coronal heating'', led by C.~Froment and P.~Antolin. MK also acknowledges the support from the Vicepresidència i Conselleria d’Innovació, Recerca i Turisme del Govern de les Illes Balears and the Fons Social Europeu 2014-2020 de les Illes Balears. PA acknowledges funding from his STFC Ernest Rutherford Fellowship (No. ST/R004285/2). DK has received funding from the S\^{e}r Cymru II scheme, part-funded by the European Regional Development Fund through the Welsh Government and from the Georgian Shota Rustaveli National Science Foundation project FR17 323. This research has made use of SunPy v1.1, an open-source and free community-developed solar data analysis Python package \citep{SunPy2020}. The Swedish 1-m Solar Telescope is operated on the island of La Palma by the Institute for Solar Physics of Stockholm University in the Spanish Observatorio del Roque de los Muchachos of the Instituto de Astrof\'\i sica de Canarias. The Institute for Solar Physics is supported by a grant for research infrastructures of national importance from the Swedish Research Council (registration number 2017-00625). Finally, the authors also acknowledge the very useful comments from the anonymous referee that helped to improve the work. 
\end{acknowledgements}

\bibliographystyle{aa}
\bibliography{citations}

\end{document}